\documentclass[journal,transmag,10pt]{IEEEtran}
\renewcommand{\baselinestretch}{1.9} \usepackage{setspace}

\usepackage{setspace} 
\usepackage{float}
\usepackage[mediumspace,mediumqspace,Grey,squaren]{SIunits}
\usepackage{caption} 
\date{}
\usepackage{amsmath,amsfonts,mathrsfs,amssymb,dsfont}
\usepackage{graphicx}
\usepackage[square,numbers,sort&compress]{natbib} 
\usepackage{booktabs}

\usepackage{bbm}
\newcommand{\et}{\hfill \blacksquare}
\newcommand{\eqdef}{\protect:=}     

\ifCLASSINFOpdf
\else
\fi

\newcommand{\subparagraph}{}

\usepackage{titlesec}
\titlespacing{\section}{0pt}{0.3\parskip}{0.3\parskip}
\titlespacing{\subsection}{0pt}{0.2\parskip}{0.2\parskip}
\titlespacing{\subsubsection}{0pt}{0.2\parskip}{0.2\parskip}

\usepackage[font=small,skip=0.5em]{caption}

\begin{document}

\title{Conditions for cell size homeostasis: A stochastic hybrid systems approach}

\author{\IEEEauthorblockN{Cesar Augusto Vargas-Garcia, Mohammad Soltani, Abhyudai Singh}


\IEEEauthorblockA{ Department of Electrical and Computer Engineering, University of Delaware, Newark, DE USA 19716.
}%

\IEEEauthorblockA{Corresponding Author: A. Singh (absingh@udel.edu)}

}


\singlespacing

\IEEEtitleabstractindextext{%
\begin{abstract}
A ubiquitous feature of living cells is their growth over time followed by division into daughter cells. How isogenic cell populations maintain size homeostasis, i.e., a narrow distribution of cell size, is an intriguing fundamental problem. We model cell size using a stochastic hybrid system, where a cell grows exponentially in size (volume) over time and probabilistic division events are triggered at discrete time intervals. Moreover, whenever division events occur, size is randomly partitioned among daughter cells. We first consider a scenario, where a timer (i.e., cell-cycle clock) that measures the time since the last division event regulates both the cellular growth and division rates. Analysis reveals that such a timer-controlled system cannot achieve size homeostasis, in the sense that, the cell-to-cell size variation grows unboundedly with time. To explore biologically meaningful mechanisms for controlling size we consider two classes of regulation: a size-dependent growth rate and a size-dependent division rate. Our results show that these strategies can provide bounded intercellular variation in cell size, and exact mathematical conditions on the form of regulation needed for size homeostasis are derived. Different known forms of size control strategies, such as, the adder and the sizer are shown to be consistent with these results. Interestingly, for timer-based division mechanisms, the mean cell size depends on the noise in the cell-cycle duration but independent of errors incurred in partitioning of volume among daughter cells. In contrast, the mean cell size decreases with increasing partitioning errors for size-based division mechanisms. 
Finally, we discuss how organisms ranging from bacteria to mammalian cells have adopted different control approaches for maintaining size homeostasis.

\end{abstract}

}

\maketitle

%
\IEEEpeerreviewmaketitle

\section{Introduction}

Stochastic hybrid systems (SHS) constitute an important mathematical modeling framework that combines continuous dynamics with discrete stochastic events.
Here we use SHS to model a universal feature of all living cells: growth in cell size (volume) over time and division into two viable progenies (daughters). A key question is how cells regulate their growth and timing of division to ensure that they do not get abnormally large (or small). This problem has ben referred to literature as \emph{size homeostasis}  and is a vigorous area of current experimental research in diverse organisms \cite{llo13,csk14,knw14,rka14,mb15,scs15,jut15,tbs15,tpp15,zw15,zar15,slj16,cu15,zbm15,blk16,aa16}. We investigate if phenomenological models of cell size dynamics based on SHS can provide insights into the control mechanisms needed for size homeostasis.

The proposed model consists of two non-negative state variables: $\boldsymbol{v}(t)$, the size of an individual cell at time $t$, and a timer $\boldsymbol{{\boldsymbol \tau}}$ that measures the time elapsed from when the cell was born (i.e.,
last cell division event). This timer can be biologically interpreted as an internal clock that regulates cell-cycle processes. Time evolution of these variables is governed by the following ordinary differential equations
\begin{equation}\label{V}
\dot{\boldsymbol v}= \alpha(\boldsymbol v,\boldsymbol {\boldsymbol \tau})\boldsymbol v, \quad \dot{\boldsymbol {\boldsymbol \tau}}=1,
\end{equation}
where the \emph{growth rate}  $\alpha(\boldsymbol v,\boldsymbol {\boldsymbol \tau}) \geq 0$ can depend on both state variables and is such that \eqref{V}
has a unique and well-defined solution $\forall t \geq 0$ (i.e., cell size does not blow up in finite time).  A constant $\alpha$ implies exponential growth over time. 

\begin{figure}[hbtp]
\centering
\includegraphics[scale=.95]{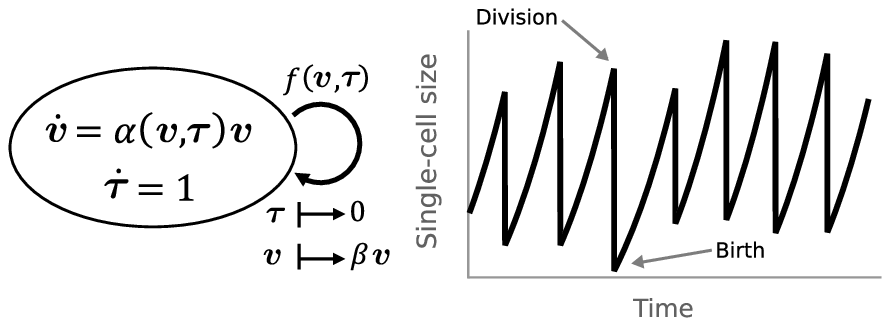}
\caption{SHS model for capturing time evolution of cell size. The size of an individual cell $\boldsymbol v(t)$ grows exponentially with growth rate $\alpha(\boldsymbol v,\boldsymbol {\boldsymbol \tau})$, where $\boldsymbol {\boldsymbol \tau}$ represents a timer that measures the time since the last division event. The arrow represents cell division events that occur with rate $f(\boldsymbol v,\boldsymbol {\boldsymbol \tau})$, which resets $\boldsymbol {\boldsymbol \tau}$ to zero and divide the size by approximately half. A sample trajectory of $\boldsymbol v(t)$ is shown with cycles of growth and division. }
\label{fig:schem}
\end{figure}

As the cell grows in size, the probability of cell division 
occurring  in the next infinitesimal time interval $(t,t+dt]$ is given by $f(\boldsymbol v,\boldsymbol {\boldsymbol \tau})dt$, where $f(\boldsymbol v,\boldsymbol {\boldsymbol \tau})$ can be interpreted as the \emph{division rate}.
Whenever a division event is triggered, the timer is reset to zero and the size is reduced to $ \beta \boldsymbol v$, where random variable  $\beta\in(0,1)$ is drawn from a beta distribution. Assuming symmetric division, $\beta$ is on average half, and its coefficient of variation ($CV_\beta$) quantifies the error in partitioning of volume between daughters. To be biologically meaningful, $\alpha(\boldsymbol v,\boldsymbol {\boldsymbol \tau})$ is a non-increasing function, while $f(\boldsymbol v,\boldsymbol {\boldsymbol \tau})$ is a non-decreasing function of its arguments.
The SHS model is illustrated in Fig. \ref{fig:schem} and incorporates two key noise sources: randomness in partitioning and timing of division. Next, we explore conditions for size homeostasis, in the sense that, the mean cell size does not converge to zero, and all statistical moments of $\boldsymbol{v}$ remain bounded. 

\section{Timer-dependent growth and division}

We begin by considering a scenario, where both the growth and division rates are functions of $\boldsymbol {\boldsymbol \tau}$, but do not depend 
on $\boldsymbol v$. The SHS can be compactly written as 
\begin{equation}\label{V1}
\dot{\boldsymbol v}= \alpha(\boldsymbol {\boldsymbol \tau})\boldsymbol v, \quad \dot{\boldsymbol {\boldsymbol \tau}}=1,
\end{equation}
with reset maps
\begin{equation}\label{V2}
\boldsymbol v\mapsto \beta \boldsymbol v, \quad \boldsymbol {\boldsymbol \tau}\mapsto 0
\end{equation}
that are activated at the time of division. The timer-controlled division rate $f(\boldsymbol {\boldsymbol \tau})$
can be interpreted as a ``hazard function" \cite{Ross20109}. Let $T_1$, $T_2$, $\ldots$ denote independent and identically distributed (i.i.d.) random variables that represent the time interval between two successive division events. Then, based on the above formulation, the probability density function (pdf) for $T_i$ is given by
\begin{equation}\label{V3}
T_i \sim f(x) e^{-\int_{y=0}^{x}f(y)dy}, \quad \forall\, x\geq 0  
\end{equation}
\cite{Ross20109}. Note that a constant division rate in \eqref{V3} would lead to an exponentially distributed $T_i$. For this class of models, the steady-state statistics of $\boldsymbol v$
is given by the following theorem. \\

\noindent {\bf Theorem 1}:
\emph{Consider the SHS \eqref{V1}-\eqref{V2} with timer-dependent growth and division rates. Then
\begin{align}\label{V4}
  \lim_{t \to \infty} \langle \boldsymbol v(t) \rangle =\begin{cases}
    0 &\left \langle e^{\int_{y=0}^{T_i}\alpha(y)dy} \right  \rangle<2\\
   \infty&\left \langle e^{\int_{y=0}^{T_i}\alpha(y)dy} \right  \rangle>2,
  \end{cases}
\end{align}
where the symbol $\langle \ \rangle$ is used to denote the expected value of a random variable. Moreover, 
\begin{align}\label{V51}
0<\lim_{t \to \infty} \langle \boldsymbol v(t) \rangle < \infty, \ \ \lim_{t \to \infty} \langle \boldsymbol v^2(t) \rangle = \infty 
\end{align} 
when $ \left \langle e^{\int_{y=0}^{T_i}\alpha(y)dy} \right  \rangle=2$}. $\et$\\
\\{\bf Proof of Theorem 1}: Let $\boldsymbol v_{i-1}$ denote the cell size just at the start of the $i^{th}$ cell cycle. Using \eqref{V1}, the size at the time of division in the 
$i^{th}$ cell cycle is given by
\begin{equation}\label{V52}
\boldsymbol v_{i-1}e^{\int_{y=0}^{T_i}\alpha(y)dy}.
\end{equation}
Thus, the size of the newborn cell in the next cycle is
\begin{equation}\label{V54}
\boldsymbol v_{i}=\boldsymbol v_{i-1}\boldsymbol x_i, \quad \boldsymbol x_i \eqdef \beta_ie^{\int_{y=0}^{T_i}\alpha(y)dy},
\end{equation}
where $\beta_i\in(0,1)$ are i.i.d random variables following a beta distribution and $\boldsymbol x_i$ are
i.i.d. random variables that are a function of $\beta_i$ and $T_i$. From \eqref{V54}, the mean cell size at the start of $i^{th}$ cell cycle is given by
\begin{equation}\label{V55}
\langle \boldsymbol v_{i} \rangle=\boldsymbol v_{0} \langle \boldsymbol x_{i} \rangle^i
\end{equation}
and will grow unboundedly over time if $\langle \boldsymbol x_{i} \rangle>1$, or go to zero if $\langle \boldsymbol x_{i} \rangle<1$. Using the fact that $\langle \beta_{i} \rangle=0.5$ (symmetric division of a mother cell into daughter cells), $\beta_{i}$ and $T_i$ are independent, \eqref{V4} is a straightforward consequence of \eqref{V55}.
It also follows from \eqref{V54} that
\begin{equation}\label{V56}
\langle \boldsymbol v^2_{i} \rangle={\boldsymbol v}^2_{0}  \langle \boldsymbol x^2_{i} \rangle^{i}={\boldsymbol v}^2_{0}  \langle \boldsymbol x_{i} \rangle^{2i}(1+CV^2_{\boldsymbol x_i})^i
\end{equation}
where $CV^2_{\boldsymbol x_i}$ represents the coefficient of variation squared of $\boldsymbol x_i$. When $\langle \boldsymbol x_{i} \rangle=1$ then $\langle \boldsymbol v_{i} \rangle={\boldsymbol v}_{0}$ and 
\begin{equation}\label{V56}
\langle \boldsymbol v^2_{i} \rangle={\boldsymbol  v}^2_{0}(1+CV^2_{\boldsymbol x_i})^i.
\end{equation}
Note that when the system is completely deterministic, i.e., pdfs for $T_i$ and $\beta_i$ are given by delta functions, $CV^2_{\boldsymbol x_i}=0$. However, the slightest noise in these variables will lead to 
$CV^2_{\boldsymbol x_i}>0$, in which case \eqref{V56} implies \eqref{V51}.$\et$

In summary, unless functions $\alpha(\boldsymbol {\boldsymbol \tau})$ and $f(\boldsymbol {\boldsymbol \tau})$ are chosen such that $ \left \langle e^{\int_{y=0}^{T_i}\alpha(y)dy} \right  \rangle=2$,
the mean cell size would either grow unboundedly or go extinct. Moreover, even if the mean cell size converges to a non-zero value, the statistical fluctuations in size would grow unboundedly. {\emph{Hence, size-based regulation of growth/division rates is a necessary condition for size homeostasis}}. 

\section{Size-dependent growth rate}

Recent work measuring sizes of single mammalian cells over time has reported lowering of growth rates as cells become bigger \cite{tkl09,kgo13,gkk15}. To explore the effects of such regulation, we consider a growth rate
$\alpha(\boldsymbol v,\boldsymbol {\boldsymbol \tau})$ that now depends on size. As in the previous section, timer-controlled division events occur with rate $f(\boldsymbol {\boldsymbol \tau})$ resulting in inter-division times $T_i$ given by \eqref{V3}. The following result shows that size homeostasis is possible if growth rate is appropriately bounded from below and above. \\

%
%
%
%
%
%
%

\noindent {\bf Theorem 2}:  \emph{Let the growth rate be bounded by
\begin{equation}\label{V7}
\alpha(\boldsymbol v,\boldsymbol {\boldsymbol \tau})\boldsymbol v \leq k(\boldsymbol {\boldsymbol \tau})\boldsymbol v^p, \ \ p\in[0,1), \ \ \forall \boldsymbol v \geq 0
\end{equation}
for some non-increasing function $k(\boldsymbol {\boldsymbol \tau})$. Moreover, the growth rate of a small cell is large enough such that 
\begin{equation}\label{V8}
\left \langle e^{\int_{y=0}^{T_i}\alpha_0(y)dy} \right  \rangle>2, \ \ \alpha_0({\boldsymbol {\boldsymbol \tau}}):= \lim_{\boldsymbol v \to 0}\alpha(\boldsymbol v,\boldsymbol {\boldsymbol \tau}).
\end{equation}
Then
\begin{equation}\label{V9}
0 < \lim_{t \to \infty} \langle \boldsymbol {\boldsymbol v}^l(t) \rangle <\left( \frac{l\langle k(\boldsymbol {\boldsymbol \tau}) \rangle\langle T_i \rangle}{\langle1- \beta^l \rangle} \right)^\frac{1}{1-p}
\end{equation}
where $l\in \{1,2,\dots \}$, $\langle T_i \rangle$ is the mean cell-cycle duration, and $\beta\in (0,1)$ is a random variable quantifying the error in partitioning of volume between daughters.} $\et$\\
\\{\bf Proof of Theorem 2}: Consider a newborn cell with a \emph{sufficiently small} size born at time $t=0$. Then, the mean cell size will grow in successive generation iff the second inequality in \eqref{V4} is true for $\alpha_0(\boldsymbol \tau)$, which results in \eqref{V8}.  Based on the Dynkin's formula for the SHS \eqref{V} and \eqref{V2}, the time evolution of moments is given by 
\begin{equation}
\begin{aligned}
 \frac{d\langle {\boldsymbol v}^l  \rangle}{dt}= & \left \langle  f(\boldsymbol {\boldsymbol \tau}) {\boldsymbol v}^l  \right \rangle  \left ( \langle \beta^l   \rangle -1 \right ) +l \left \langle  \alpha(\boldsymbol v,\boldsymbol {\boldsymbol \tau}){\boldsymbol v}^l   \right \rangle, \label{dynnf1}
  \end{aligned}
\end{equation}
for $l\in \{1,2,\dots \}$ \cite{hsi04}.  Using \eqref{V7},
\begin{equation}
\begin{aligned}
 \frac{d\langle {\boldsymbol v}^l  \rangle}{dt} \leq & \left \langle  f(\boldsymbol {\boldsymbol \tau}) {\boldsymbol v}^l  \right \rangle  \left ( \langle \beta^l   \rangle -1 \right ) +l \left \langle  k(\boldsymbol {\boldsymbol \tau}){\boldsymbol v}^{l-1+p}   \right \rangle. \label{dynnf11}
  \end{aligned}
\end{equation}
Note that
\begin{equation}\label{V91}
\left \langle  f(\boldsymbol {\boldsymbol \tau}) {\boldsymbol v}^l  \right \rangle = \left \langle  f(\boldsymbol {\boldsymbol \tau}) \langle {\boldsymbol v}^l | {\boldsymbol \tau} \rangle \right \rangle 
\end{equation}
where $\langle {\boldsymbol v}^l | {\boldsymbol \tau} \rangle$ is the expected value of ${\boldsymbol v}^l$ conditioned on ${\boldsymbol \tau}$. Based on the time evolution of cell size in \eqref{V}, $\langle {\boldsymbol v}^l | {\boldsymbol \tau} \rangle$ is an increasing function of ${\boldsymbol \tau}$ (cells further along in the cell cycle, have on average, larger sizes). Since $\langle {\boldsymbol v}^l | {\boldsymbol \tau} \rangle$ and $f(\boldsymbol {\boldsymbol \tau})$ are monotone non-decreasing function of ${\boldsymbol \tau}$
\begin{equation}\label{V92}
\left \langle  f(\boldsymbol {\boldsymbol \tau}) {\boldsymbol v}^l  \right \rangle \geq  \langle  f(\boldsymbol {\boldsymbol \tau}) \rangle \langle {\boldsymbol v}^l \rangle.
\end{equation}
Similarly, since $k(\boldsymbol {\boldsymbol \tau})$ is a  non-increasing function,
\begin{equation}\label{V93}
\left \langle  k(\boldsymbol {\boldsymbol \tau}){\boldsymbol v}^{l-1+p}   \right \rangle \leq  \langle  k(\boldsymbol {\boldsymbol \tau}) \rangle \langle \boldsymbol v^{l-1+p}  \rangle.
\end{equation}
Finally, using the fact that ${{l-1+p}} \leq {l}$ as $p\in[0,1)$
\begin{equation}\label{V94}
 \langle {\boldsymbol v}^{l-1+p}  \rangle = \left \langle \left ({\boldsymbol v}^l \right)^{\frac{l-1+p}{l}}  \right \rangle \leq \left \langle  {\boldsymbol v}^l   \right \rangle^{\frac{l-1+p}{l}}
\end{equation}
Using \eqref{V92}-\eqref{V94}, \eqref{dynnf11} reduces to the following inequality
\begin{equation}
\begin{aligned}
 \frac{d\langle {\boldsymbol v}^l  \rangle}{dt} \leq & \langle  f(\boldsymbol {\boldsymbol \tau}) \rangle \langle {\boldsymbol v}^l \rangle \left ( \langle \beta^l   \rangle -1 \right ) +l \langle  k(\boldsymbol {\boldsymbol \tau}) \rangle\left \langle  {\boldsymbol v}^l   \right \rangle^{\frac{l-1+p}{l}}. \label{dynnf111}
  \end{aligned}
\end{equation}
Since at steady state 
\begin{equation}\label{f}
\begin{aligned}
\left \langle  f(\boldsymbol\tau)\right \rangle =\frac{1}{\left \langle T_i \right \rangle},
  \end{aligned}
\end{equation}
 \cite{fin08}, \eqref{dynnf111} implies \eqref{V9}. $\et$\\

An extreme example of size-dependent growth is 
\begin{equation}\label{V71}
\alpha(\boldsymbol v,\boldsymbol {\boldsymbol \tau}) = \frac{k}{\boldsymbol v}, \ \ k>0
\end{equation}
which corresponds to cells growing linearly in size, as experimentally reported for some organisms \cite{cor03}. For this case,  the result below provides exact closed-form expressions for the 
first and second-order statistical moments of $\boldsymbol v$. \\

\noindent {\bf Theorem 3}: \emph {Consider the growth rate \eqref{V71} that results in the following SHS continuous dynamics
\begin{equation}\label{V68}
\dot{\boldsymbol v}= k, \quad \dot{\boldsymbol {\boldsymbol \tau}}=1.
\end{equation}Then, the steady-state mean and 
coefficient of variation squared of cell size is given by 
\begin{align}
\lim_{t \to \infty} \langle \boldsymbol v(t) \rangle &= \frac{k\langle T_i\rangle\left(3+CV^2_{T_i}\right)}{2}\label{cv_division_time1}, \\
CV^2_{\boldsymbol v} &= \frac{1}{27}+\frac{4\left(9\frac{\langle T^3_i\rangle}{\langle T_i \rangle^3}-9-6CV^2_
{T_i}-7CV^4_{T_i}\right)}{27\left(3+CV^2_{T_i}\right)^2}\nonumber\\&+\frac{16 CV^2_{\beta}}{3(3-CV^2_{\beta})(3+CV^2_{T_i})},\label{cv_division_time0}
\end{align}
where $CV^2_{T_i}$ and $CV^2_{\beta}$ denote randomness in the inter-division times ($T_i$) and partitioning errors ($\beta$), respectively, as 
quantified by their coefficient of variation squared.} 
 $\et$\\

The proof of Theorem 3 can be found in the Appendix. Interestingly, the mean cell size in \eqref{cv_division_time1} not only depends on the mean inter-division times $\langle T_i\rangle$, but also on its second-order moment $CV^2_{T_i}$. Thus, making the cell division times more random (i.e., increasing  $CV^2_{T_i}$) will also lead to larger cells on average. Similar effects of $CV^2_{T_i}$  on mean gene expression levels have recently been reported in literature  \cite{sva15,ans14}. Moreover,  \eqref{cv_division_time0} shows that
the magnitude of fluctuations in cell size ($CV^2_{\boldsymbol v}$) depend on 
$T_i$ through its moments up to order three. Note that if $CV^2_{\beta}=0$ (no partitioning errors) and $T_i=\langle T_i\rangle$ with probability one (deterministic inter-division times), then $CV^2_{\boldsymbol v}  = {1}/{27}.$
This non-zero value for $CV^2_{\boldsymbol v}$ in the limit of vanishing noise sources represent variability in size from cells being in different stages of the deterministic cell cycle. Theorem 3 decomposes $CV^2_{\boldsymbol v}$ into terms representing contributions from different noise sources. The terms from left to right in \eqref{cv_division_time0} represent contributions to $CV^2_{\boldsymbol v}$ from
i) Deterministic cell-cycle and ii) Random timing of division events and iii) Partitioning errors at the time of division. Assuming lognormally distributed $T_i$,
\begin{align}
\langle T^3_i\rangle/\langle T_i \rangle^3 = \left(1+ CV^2_{T_i}\right)^3. \label{app}
\end{align}
Substituting \eqref{app} in \eqref{cv_division_time0} and plotting $CV^2_{\boldsymbol v}$ as a function of $CV^2_{\beta}$ and $CV^2_{T_i}$, reveals that stochastic variations in cell size are more sensitive to partitioning errors as compared to noise in the inter-division times.

\begin{figure}[hbtp]
\centering
\includegraphics[scale=0.6]{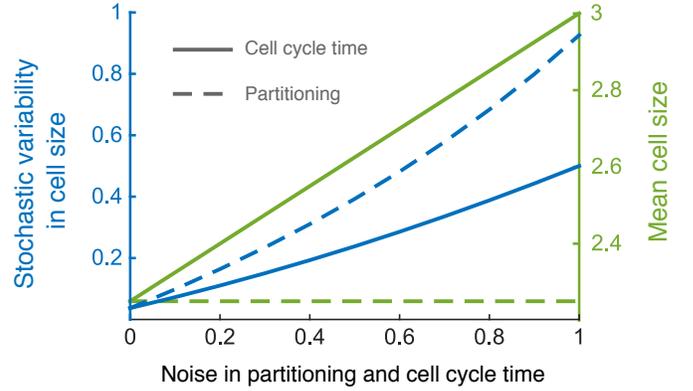}
\caption{Stochastic variation in cell size (blue) and mean cell size (green) as a function of $CV^2_{T_i}$ (noise in inter-division time) and $CV^2_{\beta}$ (error in partitioning of volume among daughters) for linear cell growth and a timer-based division mechanism. The mean cell size is dependent on $CV^2_{T_i}$ but independent of $CV^2_{\beta}$. Fluctuations in cell size increase more rapidly with $CV^2_{\beta}$ than with $CV^2_{T_i}$.}
\end{figure}

In summary, our result show that appropriate regulation of growth rate by size (as seen in mammalian cells) can be an effective mechanism for achieving size homeostasis. We next consider a different class of models where size-based regulation is at the level division rather than growth. 
 
\section{Size-dependent division rate}

In contrast to growth rate control, many organisms rely on size-dependent regulation of division rate for size homeostasis  \cite{coo06,coo13,rhk14,onl14,rat15}. To analyze this strategy, we consider the SHS continuous dynamics \eqref{V1} with a timer-dependent growth rate $\alpha(\boldsymbol \tau)$, and a division rate $f(\boldsymbol v,\boldsymbol \tau)$ that now depends on size.  The theorem below provides sufficient conditions on $f(\boldsymbol v,\boldsymbol \tau)$ for size homeostasis. \\

\noindent {\bf Theorem 4}:
\emph{ Let there exist a non-decreasing function $g(\boldsymbol \tau)$ and $p>0$ such that 
\begin{equation}\label{V81}
f(\boldsymbol v,\boldsymbol \tau) \geq g(\boldsymbol \tau) \boldsymbol v^p.
\end{equation}
Moreover, the division rate for a sufficiently small cell size $f_0(\boldsymbol \tau):= \lim_{\boldsymbol v \to 0}f(\boldsymbol v,\boldsymbol \tau)$
satisfies
\begin{equation}\label{V82}
\left \langle e^{\int_{y=0}^{T_i}\alpha_0(y)dy} \right  \rangle>2, \ \ T_i \sim f_0(x) e^{-\int_{y=0}^{x}f(_0y)dy}.
\end{equation}
Then, for the SHS given by \eqref{V1} and \eqref{V2}
\begin{equation}\label{V10}
0< \lim_{t \to \infty} \langle \boldsymbol {\boldsymbol v}^l(t) \rangle <\left(\frac{l\langle\alpha(\boldsymbol \tau)\rangle}{\langle g(\boldsymbol \tau)\rangle({1-\langle \beta^l \rangle})}\right)^\frac{l}{p}, 
 \end{equation}}
for $ l\in\{1,2,\dots\}$. $\et$ \\
\\{\bf Proof of Theorem 4}: Consider a newborn cell with a \emph{sufficiently small} size at time $t=0$. Then, based on Theorem 1, the mean size will grow over successive generations (and not go extinct) iff \eqref{V82} holds. 
Based on the Dynkin's formula for \eqref{V1}-\eqref{V2}, the time evolution of moments is given by
\begin{equation}\label{900}
\begin{aligned}
 \frac{d\langle \boldsymbol {\boldsymbol v}^l \rangle}{dt}= & \left \langle l \alpha(\boldsymbol \tau) \boldsymbol {\boldsymbol v}^l \right \rangle -  \left \langle f(\boldsymbol v,\boldsymbol \tau)\boldsymbol {\boldsymbol v}^l \right \rangle  \left  \langle 1- \beta^l   \right\rangle 
  \end{aligned}
  \end{equation}
  Using \eqref{V81}, the fact that $\alpha(\boldsymbol \tau)$ is a non-increasing function, while $g(\boldsymbol \tau)$ is a non-decreasing function,
 \begin{equation}\label{V800}
\begin{aligned}
 \frac{d\langle \boldsymbol {\boldsymbol v}^l \rangle}{dt} \leq & l\left \langle \alpha(\boldsymbol \tau) \right \rangle  \left \langle \boldsymbol {\boldsymbol v}^l \right \rangle -  \left \langle g(\boldsymbol \tau) \right \rangle  \left \langle \boldsymbol {\boldsymbol v}^{l+p} \right \rangle  \left  \langle 1- \beta^l   \right\rangle 
  \end{aligned}
  \end{equation}
  Finally, using $\left \langle \boldsymbol {\boldsymbol v}^{l+p} \right \rangle \geq \left \langle \boldsymbol {\boldsymbol v}^{l} \right \rangle^\frac{l+p}{l}$
in \eqref{V800} result in \eqref{V10} at steady state.
 $\et$\\

Next, we show that different known strategies for size-dependent regulating of inter-division times are consistent with Theorem 4. A common example of size-dependent division is the  ``sizer strategy",
 where a cell senses its size, and divides when a critical size threshold is reached  \cite{tyo86,tes12,psf14,stk15}. Such as strategy can be implemented by 
\begin{equation}\label{V810}
f(\boldsymbol v,\boldsymbol \tau) = \left(\frac{\boldsymbol v}{\bar{v}}\right)^p
\end{equation}
where ${\bar{v}}$ and $p$ are positive constant. A large enough $p$ corresponds to division events occurring when size reaches $\bar{v}$. In contrast to the sizer strategy, many bacterial species use an  ``adder strategy", where a cell divides after adding a fixed size from birth \cite{ama14,dvx15,fdm15,sra16}. In the case of exponential growth (constant growth rate $\alpha$), the adder strategy can be implemented by
\begin{equation}\label{V8100}
f(\boldsymbol v,\boldsymbol \tau) = \left(\frac{\boldsymbol v \left(1-e^{-\alpha \boldsymbol \tau }\right)}{\bar{v}}\right)^p.
\end{equation}
A large enough $p$ would correspond to cells adding a fixed size ${\bar{v}}$ between cell birth and division \cite{gvs15}. Both these division rates are consistent with the form of $f$ required for size homeostasis 
in Theorem 4. We investigate the first two moments of $\boldsymbol v$ in more detail for the sizer strategy.

Using \eqref{900} for a constant growth rate $\alpha$ and division rate \eqref{V810} results in the following moment dynamics
\begin{equation}\label{V81000}
\frac{d\left\langle \boldsymbol v^{l}\right\rangle }{dt}=l\alpha\left\langle \boldsymbol v^{l}\right\rangle -\bar{v}^{-p}\left\langle \boldsymbol v^{l+p}\right\rangle \left\langle 1-\beta^l\right\rangle .
\end{equation}
Let $\mu=\left[\left\langle \boldsymbol v\right\rangle, \left\langle \boldsymbol v^2\right\rangle \cdots\left\langle \boldsymbol v^{L}\right\rangle \right]^{T}$
be a vector of moments up to order $L$, where $L$ is the \emph{order of truncation}. Using \eqref{V81000}, the time evolution of $\mu$ can be compactly written as 
\begin{equation}\label{eq:linmn}
\frac{d\mu}{dt}=a+A\mu+C\bar{\mu}, \ \ \bar{\mu}=\left[\left\langle \boldsymbol v^{L+1}\right\rangle \cdots\left\langle \boldsymbol v^{L+p}\right\rangle \right]^{T}
\end{equation}
for some vector $a$, matrices $A$ and $C$, and $\bar{\mu}$ is the vector of higher order moments. Note that nonlinearities in the division rate lead to the well known problem of moment closure, where time evolution of $\mu$ depends on  higher-order moments $\bar{\mu}$. Moment closure techniques that express $\bar{\mu}\approx\theta\left(\mu\right)$ are typically used to solve equations of the form \eqref{eq:linmn}.
Here, we use closure schemes based on the derivative-matching technique \cite{sih07ny,sih05,sih10}, that yield analytical expressions for the steady-state moments. 
For example, $L=2$ in \eqref{eq:linmn} (second order of truncation) results in the following steady-state mean and coefficient of variation squared of cell size 
 \begin{equation}
\label{eq:meanv}
\left\langle \boldsymbol v\right\rangle \approx2^{\frac{1}{p}}\alpha^{\frac{1}{p}}\bar{v}\left(\frac{3-CV_{\beta}^{2}}{4}\right)^{\frac{p+1}{2p}}, \ CV_{\boldsymbol v}^{2}\approx\left(\frac{4}{3-CV_{\beta}^{2}}\right)^{\frac{1}{p}}-1,
\end{equation} 
respectively. 
Intriguingly, \eqref{eq:meanv} shows that the mean cell size decreases with increasing magnitude of partitioning error $CV^2_{\beta}$. While the results from \eqref{eq:meanv} are qualitatively consistent with moments obtained via Monte Carlo simulations, a much higher order of truncation is needed in \eqref{eq:linmn} to get an exact quantitative match (Fig. 3).


\begin{figure}[h!]
\centering
\includegraphics[scale=0.6]{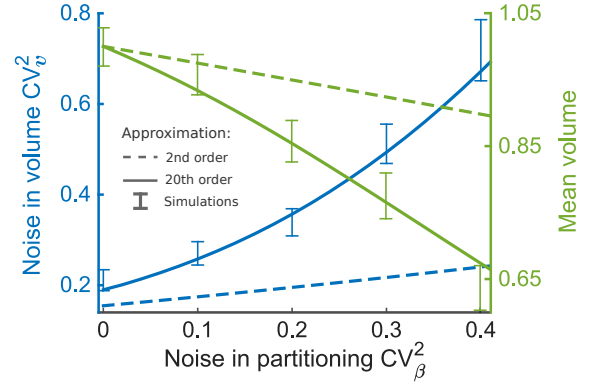}\\
\caption{Stochastic variation in cell size (blue) and mean cell size (green) as a function of $CV^2_{\beta}$ (error in partitioning of volume among daughters) for exponential cell growth and sizer-based division mechanism. The mean cell size decreases with increasing $CV^2_{\beta}$, while noise in cell size increases with it. Results are shown for a  2$^{nd}$ (dashed) and a 20$^{th}$ (solid) order moment closure truncation, and compared with moments obtained by running a large number of Monte Carlo simulations. Errors bars show $95\%$ confidence estimates.}
\label{fig:cv2v}
\end{figure}

\section{Conclusion}

Here we have used a phenomenological SHS framework to model time evolution of cell size (Fig. 1). The model is defined by three features: a growth rate $\alpha(\boldsymbol v,\boldsymbol {\boldsymbol \tau})$, a division rate $f(\boldsymbol v,\boldsymbol {\boldsymbol \tau})$, and a random variable $\beta\in (0,1)$ that determines the reduction in size when division occurs. A key assumption was that $\alpha$ and $f$ are monotone functions: with increasing size and cell-cycle progression, the growth rate decreases, and propensity to divide increases. Our main contribution was to identify sufficient conditions on $\alpha$ and $f$  that prevent size extinction and also lead to bounded moments (Theorems 2 and 4).  In essence, these conditions require the growth (division) rate to decrease (increase) with cell size in a polynomial fashion.

We also analyzed two strategies for size homeostasis: i) Linear growth in size with timer-controlled divisions and ii) Exponential growth in size with size-controlled divisions. Analysis reveals that in the former strategy, the mean cell size is independent of volume partitioning errors at the time of mitosis. In contrast, the mean cell size decreases with increasing partitioning errors for size-controlled divisions. Moreover, stochastic variations in cell size are found to be highly sensitive to partitioning errors for both strategies (Fig. 2 and 3). This suggests that cells may use mechanisms to minimize volume mismatch among daughter cells.  In summary, theoretical tools for SHS can provide fundamental understanding of regulation needed for size homeostasis. Future work will focus on coupling cell size to gene expression, and understanding how concentration of a given protein is maintained in growing cells \cite{ss16,pnb15,ShM15,mb12}.


\section*{Appendix: Proof of Theorem 3}

We prove Theorem 3 in the following steps: we use forward Kolmogorov equation to derive the equation describing the probabilistic evolution of timer $\boldsymbol\tau$. We use our derivation to calculate the probability distribution of timer and its moments. Next we use forward Kolmogorov equation again to derive the equation describing the joint probability distribution of timer $\boldsymbol\tau$ and volume $\boldsymbol{v}$ and we calculate the steady-state conditional mean volume $\overline{\langle \boldsymbol{v}\vert \boldsymbol\tau \rangle}$; we denote the steady-state mean by $\overline{\langle \ \rangle}$. Lastly we uncondition $\overline{\langle \boldsymbol{v}\vert \boldsymbol\tau \rangle}$ to obtain $\overline{\langle \boldsymbol{v} \rangle}$. We repeat the same steps for deriving $\overline{\langle \boldsymbol{v}^2 \rangle}$.

\subsection{The probability distribution of the timer $\boldsymbol \tau$}
Using forward Kolmogorov equation for stochastic hybrid systems \cite{kam92}, the probability distribution of timer $p(\tau)$ at steady-state is described through 
\begin{equation}
\frac{\partial p(\tau)}{\partial \tau} = - f(\tau)p(\tau), \ \ \tau>0, \label{cme tau}
\end{equation}
where $\tau$ is the dummy variable for $\boldsymbol{\tau}$.
We start our analysis by taking integral from both sides of \eqref{cme tau}
\begin{equation}
\frac{\partial p(\tau)}{\partial \tau } = f(\tau)p(\tau)\Rightarrow p(\tau) = p_0 {\rm e}^{-\int_0^{\tau} f(y) d y},
\end{equation}
where $p_0$ is a normalization constant. It can be shown that $p_0=\overline{\langle f(\boldsymbol \tau) \rangle}$ 
\begin{equation}\begin{aligned}
&\overline{\langle f(\boldsymbol \tau) \rangle} = p_0 \int_0^{\infty}f(\tau){\rm e}^{-\int_0^\tau f(y) d y} d \tau \\ &\Rightarrow \overline{\langle f(\boldsymbol \tau) \rangle} = p_0 \left(- {\rm e}^{-\int_0^\tau f(y) d y}\right)_0^\infty \Rightarrow\overline{\langle f(\boldsymbol \tau) \rangle} = p_0.
\end{aligned}
\end{equation}
Hence $p(\tau)$ can be written as
\begin{equation}
p(\tau) = \frac{1}{\langle T_i\rangle} {\rm e}^{-\int_0^{\tau} f(y) d y}. \label{prob. tau}
\end{equation}

Moreover moments at steady state of $\boldsymbol \tau $ can be calculated from the probability distribution
\begin{equation}
\overline{\langle {\boldsymbol \tau }  \rangle} =\frac{1}{\langle T_i\rangle} \int_0^{\infty}  \tau {\rm e}^{-\int_0^{\tau} f(y) d y} d \tau . \label{mean tau}
\end{equation} 
Note that from equation (4) in the main text we can calculate the second order moment $\langle T_i^2 \rangle $ as
\begin{equation}
{\langle T_i^2 \rangle } = \int_0^{\infty} x^2 f(x){\rm e}^{-\int_0^{x} f(y) d y} d x, 
\end{equation}
integrating by parts results in 
\begin{equation}
\langle T_i^2 \rangle =2\int_0^{\infty}x{\rm e}^{-\int_0^{x} f(y) d y} d x. \label{tau}
\end{equation}
Hence by a change of variables in \eqref{tau} and using \eqref{mean tau} we have 
\begin{equation}
\overline{\langle {\boldsymbol \tau }  \rangle} = \frac{{\langle T_i^2 \rangle }}{2{\langle T_i \rangle }}. \label{mean of tau}
\end{equation}
Using similar analysis results in 
\begin{equation}
\overline{\langle {\boldsymbol \tau^2 }  \rangle} = \frac{{\langle T_i^3 \rangle }}{3{\langle T_i \rangle }}.\label{second moment of tau}
\end{equation}

\subsection{Mean of cell volume $\overline{\langle {\boldsymbol{v}}  \rangle }$}

From the forward Kolmogorov equation,  the joint probability distribution of the timer and the volume $p(\tau, v)$ at steady-state is given by
\begin{equation}\begin{aligned}
\frac{\partial p(\tau, {v})}{\partial \tau} + \frac{\partial }{\partial {v} }\left(  k  p(\tau,{v})\right)  =
-f(\tau) p(\tau, {v}), \ \ \tau>0.
 \end{aligned}
 \label{cme}
\end{equation}  
The conditional mean volume $\overline{\langle {\boldsymbol{v}} \vert \boldsymbol\tau  \rangle}$ can be written as
\begin{equation}
\begin{aligned}
\overline{\langle {\boldsymbol{v}} \vert \boldsymbol\tau  \rangle}\equiv\overline{\langle {\boldsymbol{v}} \vert \boldsymbol\tau =\tau \rangle}=&\frac{1}{p(\tau )}\int_{0}^{+\infty} v p(\tau, v)dv.
\label{cond. x'}
\end{aligned}
\end{equation}
Taking derivative with respect to $\tau$ from \eqref{cond. x'} results in
\begin{equation}
\begin{aligned}
\frac{\partial \overline{\langle {\boldsymbol{v}} \vert \boldsymbol\tau  \rangle }}{\partial \tau } =& -\frac{\frac{\partial p (\tau)}{\partial \tau}}{p^2(\tau )} \int_{0}^{+\infty}v  p(\tau, v)dv\\
&+\frac{1}{p(\tau )}\int_{0}^{+\infty} v \frac{\partial p(\tau, v)}{\partial \tau} 
dv .
\label{derivate00}
\end{aligned}
\end{equation}

To calculate $\frac{\partial \overline{\langle {\boldsymbol{v}} \vert \boldsymbol\tau  \rangle }}{\partial \tau } $ we need expressions for $\frac{\partial p(\tau, v)}{\partial \tau}$ and $\frac{\partial p(\tau)}{\partial \tau}$.
Substituting these expressions from \eqref{cme} and \eqref{cme tau} in \eqref{derivate00} and performing some algebraic manipulations gives
\begin{equation}\begin{aligned}
\frac{\partial  \overline{\langle \boldsymbol{v} \vert \boldsymbol\tau \rangle} }{\partial \tau} =k
  \end{aligned}
 \label{conditional xx00}
\end{equation}
Hence the mean volume given the timer is given by
\begin{equation}\begin{aligned}
& \overline{\langle \boldsymbol{v} \vert \boldsymbol\tau \rangle} = k\tau +\overline{\langle \boldsymbol{v} \vert \boldsymbol\tau =0 \rangle}.
  \end{aligned}
 \label{conditional x'}
\end{equation} 

To calculate $\overline{\langle \boldsymbol{v} \vert \boldsymbol\tau=0 \rangle}$ we use equation (3) in the main text. Note that in the time of division $\boldsymbol\tau \mapsto 0$, hence in the time of division $\boldsymbol{v} \vert \boldsymbol\tau=T_i \mapsto \boldsymbol{v} \vert \boldsymbol\tau=0$. Given the fact that the volume before and after  division are related via (3), $\boldsymbol{v} \vert \boldsymbol\tau=0$ is equal to $\beta  \ \boldsymbol{v} \vert \boldsymbol\tau=T_i, \ i=\{1,2,\ldots\} $. Hence the mean volume after division is related to the mean volume right before division as
\begin{equation}
\begin{aligned}
\langle \boldsymbol{v} \vert \boldsymbol\tau=0 \rangle = \langle \beta \rangle \boldsymbol \langle \boldsymbol{v}\vert \boldsymbol\tau =T_i \rangle , \ i=\{1,2,\ldots\}.
\end{aligned} 
\end{equation}
Thus the mean volume after divisions can be written as
\begin{equation}
\begin{aligned}
\overline{ \langle \boldsymbol{v} \vert \boldsymbol\tau=0 \rangle }= \langle \beta \rangle \overline{ \langle \boldsymbol{v}\vert \boldsymbol\tau = \langle T_i \rangle  \rangle }.\label{initial v} 
\end{aligned} 
\end{equation}

By substituting \eqref{initial v} in \eqref{conditional x'} we have
\begin{equation}
\overline{\langle \boldsymbol{v} \vert \boldsymbol\tau=0 \rangle}= \frac{\langle \beta \rangle }{1-\langle \beta \rangle} \langle T_i\rangle .
\end{equation}
Thus equation \eqref{conditional x'} can be written as
\begin{equation}
\overline{\langle \boldsymbol{v} \vert \boldsymbol\tau \rangle}= k \tau +k \frac{\langle \beta \rangle }{1-\langle \beta \rangle} \langle T_i\rangle . \label{mean000}
\end{equation}

The mean volume $\overline{\langle {\boldsymbol v}  \rangle}$ can be derived by multiplying $p(\tau)$ from \eqref{prob. tau} to equation \eqref{mean000}, taking integral from both sides, and using \eqref{mean of tau}

\begin{equation}
\begin{aligned}
& \overline{\langle \boldsymbol{v}  \rangle} = k  \frac{\langle T_i^2 \rangle }{2\langle T_i\rangle } 
 +k   \frac{\langle \beta \rangle }{1-\langle \beta \rangle} \langle T_i\rangle \Rightarrow \\
& \overline{\langle \boldsymbol{v}  \rangle} =  k  \langle T_i\rangle  \frac{ 1+\langle\beta \rangle }{2(1-\langle \beta \rangle)}+\frac{k  \langle T_i\rangle CV^2_{T_i} }{2}.
 \end{aligned}
 \label{v}
\end{equation} 

\subsection{The second order moment of the volume $\overline{\langle \boldsymbol{v}^2 \rangle }$}
The sketch of the proof for the second order moment is similar to the mean of the volume. We start by deriving the conditional moment $\overline{\langle {\boldsymbol{v}^2} \vert \boldsymbol\tau  \rangle}$, and then we uncondition it to calculate $\overline{\langle {\boldsymbol{v}^2}  \rangle}$; the conditional second order moment of the volume is defined as
\begin{equation}
\begin{aligned}
\overline{\langle {\boldsymbol{v}^2} \vert \boldsymbol\tau  \rangle }  \equiv \frac{1}{p(\tau )}\int_{0}^{+\infty} v^2 p(\tau, v)dv.
\label{cond. v2}
\end{aligned}
\end{equation}
Taking derivative with respect to $\tau$ from \eqref{cond. v2} results in 
\begin{equation}
\begin{aligned}
\frac{\partial \overline{\langle {\boldsymbol{v}^2} \vert \boldsymbol\tau  \rangle } }{\partial \tau } =-\frac{\frac{\partial p (\tau)}{\partial \tau}}{p^2(\tau )} \int_{0}^{+\infty}v^2  p(\tau, v)dv\\ +\frac{1}{p(\tau )}\int_{0}^{+\infty} v^2 \frac{\partial p(\tau, v)}{\partial \tau} 
dv .
\label{derivate v2}
\end{aligned}
\end{equation}
Substituting \eqref{cme} and \eqref{cme tau} in \eqref{derivate v2} yields
\begin{equation}\begin{aligned}
&\frac{ \overline{\langle \boldsymbol{v}^2 \vert \boldsymbol\tau  \rangle } }{\partial \tau} =2 k  \overline{\langle \boldsymbol{v} \vert \boldsymbol\tau  \rangle }\Rightarrow \frac{ \overline{\langle \boldsymbol{v}^2 \vert \boldsymbol\tau  \rangle }}{\partial \tau}  =  2 k ^2 \tau +2k ^2  \frac{\langle \beta \rangle }{1-\langle \beta \rangle} \langle T_i\rangle  .
  \end{aligned}
 \label{derivative v2}
\end{equation} 
Thus $\overline{\langle {\boldsymbol{v}^2} \vert \boldsymbol\tau \rangle}$ can be written as
\begin{equation}
\overline{\langle {\boldsymbol{v}^2} \vert \boldsymbol\tau \rangle}=   k ^2 \tau^2 +2k ^2  \frac{\langle \beta \rangle }{1-\langle \beta \rangle} \langle T_i\rangle \tau +  \overline{\langle {\boldsymbol v^2} \vert \boldsymbol\tau=0 \rangle}.\label{conditial v2} 
\end{equation}
In order to calculate $\overline{\langle {\boldsymbol v^2} \vert \boldsymbol\tau=0 \rangle}$ we use equation (3) in the main article
\begin{equation}\begin{aligned}
\overline{\langle {\boldsymbol v^2} \vert \boldsymbol\tau=0 \rangle}    =   \langle \beta^2 \rangle \overline{\langle {\boldsymbol v^2} \vert \boldsymbol\tau=\langle T_i \rangle \rangle} . \end{aligned} 
\label{initial v22}
\end{equation}
Using \eqref{initial v22} in \eqref{conditial v2} results 
\begin{equation}
\overline{\langle {\boldsymbol v^2} \vert \boldsymbol\tau=0 \rangle} =\frac{ k ^2 \langle T_i\rangle^2 \langle \beta \rangle ^2 (1+CV^2_\beta) (\frac{1+\langle \beta \rangle}{1-\langle \beta \rangle}+CV^2_{T_i})}{1-\langle \beta \rangle^2 (1+CV^2_\beta)} .\label{initial v2}
\end{equation}

Hence $\overline{\langle {\boldsymbol v^2}  \rangle}$ can be derived by unconditioning \eqref{conditial v2}, and using \eqref{second moment of tau}
\begin{equation}
 \overline{\langle {\boldsymbol  v^2}  \rangle} = k ^2\frac{\langle T_i^3 \rangle}{3\langle T_i\rangle} +k ^2  \frac{\langle \beta \rangle }{1-\langle \beta \rangle}  \langle T_i^2 \rangle + \overline{\langle {\boldsymbol v^2} \vert \boldsymbol \tau=0 \rangle},
 \label{v2'}
\end{equation} 
in which initial condition is given by \eqref{initial v2}. By having the second order moment one can calculate the noise by deriving $CV^2_{\boldsymbol v}$; for example by selecting $\langle \beta \rangle =\frac{1}{2}$, noise is quantified as equation (28) in the main text.


\end{document}